# Global perspectives on the energy landscapes of liquids, supercooled liquids, and glassy systems: Geodesic pathways through the potential energy landscape

Chengju Wang and Richard M. Stratt

Department of Chemistry
Brown University
Providence, RI 02912

PACS numbers: 64.70.Pf, 66.10.-x, 61.20.Ja, 61.43.Fs

**Abstract**

How useful it is to think about the potential energy landscape of a complex many-body system depends in large measure on how direct the connection is to the system's dynamics. In this paper we show that, within what we call the potential energy landscape ensemble, it is possible to make direct connections between the geometry of the landscape and the long-time dynamical behaviors of systems such as supercooled liquids. We show, in particular, that the onset of slow dynamics in such systems is governed directly by the lengths of their *geodesics* — the shortest paths through their landscapes within the special ensemble. The more convoluted and labyrinthine these geodesics are, the slower that dynamics is. Geodesics in the landscape ensemble have sufficiently well-defined characteristics that is straightforward to search for them numerically, a point we illustrate by computing the geodesic lengths for an ordinary atomic liquid and a binary glass-forming atomic mixture. We find that the temperature dependence of the diffusion constants of these systems, including the precipitous drop as the glass-forming system approaches its mode-coupling transition, is predicted quantitatively by the growth of the geodesic path lengths.

# I. Introduction

One of the more seductive analogies in the study of slow condensed-phase dynamics is the notion of a "reaction path."[1,2] The often unstated assumption is that wherever one travels in the configuration space of a system, the dominant configurations are somehow associated with the lowest possible potential energies in the immediate vicinity. Pathways through the system are presumed to start near some local minimum (a "reactant") and head towards another local minimum (a "product") with the rate largely determined by the need to make the highest energy configuration along the intervening route (the "transition state") a saddle point so that the whole process can have as low an activation energy as possible.[3,4]

Such notions are much easier to visualize with the few-degree-of-freedom problems encountered in chemical reaction dynamics than they are in the enormously larger dimensionality required to span all of the possibilities of a many-body system. Nonetheless this language is as mathematically well defined for motion over the *potential energy landscape* of a liquid[5-7] as it is for the potential surfaces relevant to chemical reactions. In the language of the field, the local minima are the inherent structures of the landscape and the configurations associated with each such inherent structure are those within its basin of attraction.[8-22] The rearrangement of one hydrogen-bonding configuration in water into another, for example, can be regarded as a familiar activated event in this picture.[23]

Buoyed by this analogy there has been an enormous, and rather successful, attempt to predict the rates and study the mechanism of slow condensed-phase processes by searching for transition paths.[23-26] But, not every slow process fits so neatly into this paradigm. There is no particular physical reason to think of either the slow diffusion of supercooled liquids[27,28] or the slow solvent relaxation following certain kinds of solute excitation[29] as necessarily being caused by direct basin-to-basin transitions on the energy landscape. If a system is willing to travel far enough to find the right routes, it might have enough thermal energy to skate well above some fraction of the variations in the potential surface topography. The dynamics could then still be quite slow, but that slowness would have its origin not in the size of the potential energy barriers it had to surmount, but in the lengthy and convoluted byways that it had to navigate to avoid those barriers. In the high-dimensional world of a liquid, it is easy to imagine how such entropic (as opposed to energetic) impediments could become increasingly important.[30-32]

So how can one understand the molecular origins of slow dynamics in liquids within the landscape paradigm if one does not want to be tied to the chemical reaction analogy? The idea we want to pursue here is that the geometry of the potential energy landscape is indeed important, but it is *the extended pathways through the landscape that dictate the slow dynamics*, rather than any special attributes of the landscape's critical points (its minima and saddles).[33,34] That is, it is the global and not the local features of the landscape that matter. When the dynamics becomes sufficiently slow, this picture suggests that the dominant pathways through the configuration space are those that somehow manage to be the most efficient.[35-39] The goal of this paper is to talk about what constitutes an optimally efficient path in a potential energy landscape and to explain how we can find these special pathways in liquids.

The general problem of finding paths through the high-dimensional space intrinsic to condensed-matter problems has been the subject of significant amount of effort over the years,[40-51] but the issue of long-distance travel over rugged potential energy landscapes has not received nearly as much attention. We therefore need to strike out in a somewhat different direction.

The approach we shall take here is to consider what paths through a landscape would look like if the system were restricted to lie within a different ensemble than is commonly used, the *potential energy landscape ensemble*.[52] This ensemble, which we introduced in the companion paper to this one, consists of all N-molecule configurations of the liquid $\mathbf{R} = \{\mathbf{r}_1, \ldots, \mathbf{r}_N\}$ (with $\mathbf{r}_j$ the coordinates of the j-th molecule) whose potential energy $V(\mathbf{R})$ is less than or equal to some landscape energy $E_L$. As one can see from Figure 1, this ensemble has the interesting property of not having any potential barriers. Standard entropy maximization arguments imply that all of the allowed configurations in the ensemble ($V(\mathbf{R}) < E_L$) are equally probable, and the definition of the ensemble implies that all of the remaining configurations ($V(\mathbf{R}) > E_L$) are forbidden.[52] There are thus no locations on the potential surface that are possible but unlikely – and there are no extended waits for energy fluctuations large enough to reach such locations.

The complete absence of activated events makes this kind of picture certainly seems to be at odds with more conventional perspectives. In fact, results in the landscape ensemble are not even obtained as functions of the experimentally relevant variable temperature T; they show up as functions of landscape energy. However, the equivalence of ensembles in thermodynamic limit[53] suggests that these two pictures are simply complementary and that the variables T and $E_L$ should be thermodynamically equivalent. Indeed, our previous work[52] has shown that given a landscape energy $E_L$, we can compute the equivalent temperature T directly from $P(E; E_L)$, the probability distribution of potential energies E

$$P(E; E_L) = \langle \delta(E - V(\mathbf{R})) \rangle_{E_L}$$

$$= \frac{\int d\mathbf{R}\, \delta(E - V(\mathbf{R}))\, \theta(E_L - V(\mathbf{R}))}{\int d\mathbf{R}\, \theta(E_L - V(\mathbf{R}))} . \quad (1.1)$$

$$(k_B T)^{-1} = \lim_{E \to E_L^-} \frac{\partial \ln P(E; E_L)}{\partial E} . \quad (1.2)$$

In these expressions $k_B$ is Boltzmann's constant and $\theta(x) = (1, x \geq 0; 0, x < 0)$. Note that our formula for $P(E; E_L)$ explicitly uses the result that all of the allowed configurations $\mathbf{R}$ in the landscape ensemble are equally likely.[52]

So how can we go about finding the optimum paths in this ensemble? The presence of forbidden regions with $V(\mathbf{R}) > E_L$ means that the landscape is littered with a variety of impenetrable obstacles. On the other hand, both model calculations and numerical calculations on realistic liquids make it clear that the probability distribution $P(E; E_L)$ is sharply peaked near $E = E_L$ for macroscopic systems[52] – implying that the paths should remain close to constant-potential-energy contours of the landscape. Our problem then is to study what is effectively force-free diffusive motion around a series of hard obstacles. But when the dynamics is slow, we will show that this motion, in turn, is dominated by the very shortest accessible paths, the *geodesics* of the potential energy landscape. Thus what we really need to do is to learn how to find these geodesics for a liquid and to study them as a function of landscape energy.

The remainder of this paper is organized as follows: Section II discusses precisely why the optimum paths in our ensemble are, in fact, geodesics, and describes the geometry one should expect from such paths. It goes on to demonstrate that knowing the lengths of these geodesics allows us to determine diffusion constants. Section III shows how one can turn what we know about the geometry of our landscape geodesics into an explicit algorithm for finding geodesics in liquids. After specifying our model systems and some numerical details in Sec. IV, we then illustrate our basic approach in Sec. V by locating geodesic paths in both an atomic liquid and a standard model for a glass-forming liquid. We show, in particular, how the growth of the geodesic path lengths as one lowers the landscape energy accurately predicts the decrease in diffusion constants that one see with decreasing temperature. We conclude in Sec. VI with a few general remarks.

## II. Geodesic paths in the potential energy landscape ensemble

### A. The nature of the optimum paths

If one is diffusing very slowly at a constant potential energy, albeit with very irregularly boundary conditions, what are the most important paths that the system follows? A formal way to tackle this problem is to write the solution of the Fokker-Planck equation as a path integral.[54-56] The probability density for propagating from some configuration $\mathbf{R}_0$ to some configuration $\mathbf{R}$ in a time t is given by the Green's function

$$G(\mathbf{R}_0 \to \mathbf{R} | t) = \int D\,\mathbf{R}(\tau) \, \exp\left\{-\frac{1}{4D} \int \left(\frac{d\mathbf{R}}{d\tau}\right)^2 d\tau\right\} \quad . \tag{2.1}$$

As the diffusion constant $D \to 0$, the paths that dominate are going to be the paths that minimize the path-integral action.[38,54-57] That is, the key paths are the ones that obey the principle of classical-mechanical least action.

$$\delta S[\mathbf{R}(\tau)] = 0 \quad , \quad S[\mathbf{R}(\tau)] = \int (2\,T)\,d\tau \quad , \tag{2.2}$$

with T the kinetic energy (as opposed to the ones obeying Hamilton's principle, which is appropriate for a variable energy but fixed time).[58-61]

But for a general set of coordinates this observation also means that the constant-potential-energy paths must also minimize the "kinematic path length" $\ell$

$$\ell = \int ds \tag{2.3a}$$

$$(ds)^2 = \sum_{\mu\nu} g_{\mu\nu}\, dx^\mu\, dx^\nu = 2T\, (d\tau)^2 \tag{2.3b}$$

$$T = \tfrac{1}{2} \sum_{\mu\nu} g_{\mu\nu} \frac{dx^\mu}{d\tau} \frac{dx^\nu}{d\tau} \quad , \tag{2.3c}$$

where $\mu$ and $\nu$ label all of the degrees of freedom. The metric defined by the tensor $g_{\mu\nu}$ is called the kinematical metric.[62] Hence, the paths that we want, the paths with the shortest kinematic path length, are precisely the geodesics for this metric.

A few examples: For an atomic liquid with all of the particles having the same mass m:

$$g_{\mu\nu} = m\, \delta_{\mu\nu} \quad , \quad T = \tfrac{1}{2} \sum_\mu m \left(\frac{dx^\mu}{dt}\right)^2 \quad , \tag{2.4}$$

in Cartesian coordinates, making the kinematic path length proportional to the actual path length

$$\ell = \int ds = m^{\tfrac{1}{2}} \int \sqrt{\sum_\mu (dx^\mu)^2} \quad . \tag{2.5}$$

so the key paths are literally the shortest paths. (That is, the paths obey Fermat's principle.)[58-60] It is easy to generalize these results to an atomic liquid mixture in which some of the atoms have different masses than others:

$$g_{\mu\nu} = m_\mu\, \delta_{\mu\nu} \quad , \quad T = \tfrac{1}{2} \sum_\mu m_\mu \left(\frac{dx^\mu}{dt}\right)^2 . \tag{2.6}$$

where $m_\mu$ is the mass associated with the $\mu$-th degree of freedom. The kinematic path length is now a mass weighted version of the actual path length

$$\ell = \int ds = \int \sqrt{\sum_\mu m_\mu (dx^\mu)^2} \quad , \tag{2.7}$$

(which the reader will recognize as the standard mass weighting used in normal-mode analysis).[63] Minimizing this path length is obviously best accomplished by having the heaviest atoms move the least. Significantly for our future applications, we can also incorporate molecular liquids within the same formalism. For rigid molecules, the coordinates now include the (non-Cartesian) Euler angles, but the same basic equations apply.[62,64] Quite generally, *the optimum path is the geodesic (mimimum "length" path) for the kinematical metric*.

In the absence of constraints on where the system can go, this geodesic is just the classical trajectory for the free (constant-potential-energy) system. (The role of time, though, can be played by any variable that traces the progress from start to finish. Here we choose "time" to be a dimensionless quantity $\tau$ that we set equal to 0 at the start and 1 and the end.) Formally, the requirement that the kinematic length be stationary implies a stationary action

$$\frac{\delta \ell[\mathbf{x}(\tau)]}{\delta \mathbf{x}(\tau)} = 0 \Rightarrow \frac{\delta S[\mathbf{x}(\tau)]}{\delta \mathbf{x}(\tau)} = 0$$

which, in turn, predicts that the required geodesics are free-particle classical trajectories.[58-60] For the atomic liquid case (with any choice of masses), that trajectory is just straight-line motion with the same "velocity" for each coordinate.

$$x^\mu(\tau) = x^\mu(0) + \tau \left[ x^\mu(1) - x^\mu(0) \right] , \quad 0 \leq \tau \leq 1 , \tag{2.8}$$

whereas for rigid molecules without center-of-mass motion, the trajectory is described by Euler's equations for the free rigid-body rotation.[62,64]

What do our constraints $V(\mathbf{R}) \leq E_L$ do to these paths? In typical statistical mechanical applications the constraints are equalities rather than inequalities – which one handles using Lagrange multipliers. However, this idea can easily be generalized to allow for inequality constraints by applying the Kuhn-Tucker theorem.[65] The theorem says that the constrained stationary solutions consists of segments that either (1) satisfy the constraint as a strict inequality ($V(\mathbf{R}) < E_L$), in which case the answer is the same as the one computed with the unconstrained stationary condition (Eq. (2.8) for example), or (2) satisfy the constraint as an equality ($V(\mathbf{R}) = E_L$), making the answer the same as the Lagrange-multiplier-plus-stationary-condition result. But the former is just ordinary free-particle classical motion, and the latter is a requirement that the system move only along a potential-energy-equals-total-energy boundary. For the atomic liquid cases, that means that *the geodesic path consists of motion along the boundaries of allowed regions joined by straight-line segments* (Fig. 2a).[66]

This last condition is actually very powerful. Even without carrying out the explicit computation outlined in (2), simply knowing the form of the answer gives us sufficient guidance to devise an algorithm for locating possible geodesic paths in a liquid; no small thing given the high dimensionality of the configuration space.[67,68] However, by itself, this condition is not enough to specify all of the details of our geodesic paths. As illustrated in Fig. 2b, our requirement does not specify when the optimum path lies along the boundary and when it travels between boundaries. Nonetheless, as we show in Sec. III, it is straightforward to come up with algorithms that produce paths falling in what we might call the "Kuhn-Tucker class" (satisfying our general condition), which we can then optimize with a series of local moves.

**B. Some literature connections**

Our argument for the dominance of landscape geodesics in slow-diffusive motion has some strong parallels with an analogous argument that has been advanced for a geometric interpretation of facilitated kinetics.[57] There the idea is that kinetic constraints effectively select a limited subspace from the whole configuration space, effectively imposing a complicated metric on the dynamics. In the limit of low temperature ($D \to 0$) this metric creates much more of a burden than any created by the specifics of the potential surface, so the dynamics is determined entirely by geometry – meaning that the optimum dynamics follows the geodesic path. But while this picture is formally very similar to ours, it is not as easily connected with the underlying microscopics. The space in which facilitated-kinetics takes place is assumed to be some coarse-grained version of the molecular degrees of freedom[69,70] so there is, at best, an indirect route from the real potential surface to the facilitation rules that are responsible for the slow dynamics. Indeed, the success of facilitated kinetics pictures has been taken as evidence that a potential-energy-landscape perspective will probably not be of much help in understanding the onset of slow dynamics.[71]

Within our ensemble, though, the connections between microscopically derived potential surfaces and the geometry of the efficient paths are quite straightforward. In that sense our work is closer in spirit to a number of the published variational approaches to finding the most efficient paths through complex landscapes. In explicitly minimizing an action, our formalism seems to have connections both to that of Onsager and Machlup[55,72,73] and to more recent "action-derived molecular dynamics".[45-51] However the fact that our paths here are limited to configuration space rather than phase space has some significant consequences. Not only are our paths much simpler than the real classical dynamical paths found by looking at variations in the action, the paths we are looking for are more likely to be genuine minima rather than simply stationary points.

The significance here is that the non-minimum character of the action frequently poses difficulties in numerically implementing variational formulations of classical mechanics.[46-48,74] Even the one-dimensional harmonic oscillator faces this issue when formulated in the standard fashion.[74] In the potential energy landscape, by contrast, the harmonic-oscillator geodesic path length is always a minimum for any path length, in any number of dimensions; the potential surface of a multi-dimensional harmonic oscillator is a bowl, making the mimimum length path the direct line between the initial and final points.

These differences not withstanding, we should mention at least one important commonality between our landscape geodesic and many of the literature approaches to path finding: an emphasis on paths defined by double-ended boundary conditions.[73-77] Although in our approach we are not looking for minimum energy (or minimum free energy) path usually being sought after, a number of path-finding problems have the same basic structure of needing to find some sort of optimum route between two specified locations in coordinate space. As such, the experience developed in the literature may still prove a valuable technical resource.

**C. An application: geodesic path lengths and diffusion constants**

One application of our geodesic results that we can make almost immediately is to the calculation of the diffusion constant of a liquid. By our arguments in Sec. II.A, the probability density for our liquid to diffuse is given by the Green's function for free diffusion around infinitely hard objects. But the path-integral description of such a Green's function has been known for some time. Lieb pointed out that finding the exchange second virial coefficient for a hard-sphere Fermi or Bose fluid is equivalent to studying the diffusion between two points with an intervening impenetrable spherical obstacle.[78,79] The resulting action depends simply on the length of the path taken between the end points, so the dominant contribution comes from the shortest such path.

$$G(\mathbf{R} \to \mathbf{R}' \mid t) \sim (4\pi D_0 t)^{-d/2} e^{-g^2/(4D_0 t)}, \qquad (2.9)$$

where g is the length of this geodesic path, $D_0$ is what diffusion constant would be in the absence of obstacles, and d is the spatial dimension.

Although one could, in principle get contributions from many different contributing paths, when the overall diffusion is slow, the same reasoning works should work for diffusion around a *set* of infinitely hard obstacles, leading to the same mathematical expression for the dominant contribution to our Green's function in terms of the geodesic path length. On the other hand, since the motion we are describing can also be regarded as free diffusion in the d = 3N dimensional configuration space, the same Green's function can be written as

$$G(\mathbf{R} \to \mathbf{R}' \mid t) = (4\pi D t)^{-d/2} e^{-(\mathbf{R}' - \mathbf{R})^2/(4Dt)}, \qquad (2.10)$$

where D is the phenomenological (experimental) diffusion constant. Comparing the two expressions implies that we should be able to predict the behavior of experimental diffusion constants from the geometry of the landscape alone, just by computing the ratio of the Euclidean and geodesic distances, $\Delta R = |\mathbf{R}' - \mathbf{R}|$ and g

$$D = \lim_{\Delta R \to \infty} D_0 \; \overline{(\Delta R/g)^2} \;, \tag{2.11}$$

with the overbar representing an average over the possible end points $\mathbf{R}$ and $\mathbf{R}'$.

Equation (2.11) is one of our key findings. It directly embodies the idea that diffusion slows down precisely because the lengths of the shortest available paths g are beginning to grow. It also lays the foundation for practical numerical calculations in that it predicts that diffusion constant should depend only on the ratio of the geodesic path length to the Euclidean (direct) distance $\Delta R$, not on the individual endpoints or on the actual distance between them.

To carry out a real calculation, we will need to make some assumption about the "obstacle-free" diffusion constant $D_0$. Physically, this quantity represents the high-temperature limit of the diffusion, so we can simply assume an expression of the form

$$D_0 = \mu T, \tag{2.12}$$

with $\mu$ an unknown (but temperature-independent) bare mobility constant. We shall see in Sec. V how well these formulas represent the actual slowing down of liquid motion as one lowers the temperature.

### III. Finding geodesic paths in liquids

Since we know that basic form that geodesic paths take in our landscape ensemble, our first task to devise an algorithm that can find paths with the required (Kuhn-Tucker-theorem-satisfying) behavior. The particular algorithm we use in this paper to find these paths is portrayed schematically in Fig. 3. Given a pair of initial and final configurations $\mathbf{R_i}$ and $\mathbf{R_f}$, we attempt incremental steps in the direction of the final destination $\mathbf{R_f}$. If step length is $\delta R$ and $\mathbf{R}^{(t)}$ is the t-th step along the path, the trial location for the (t+1)-st step is

$$\mathbf{R}_0^{(t+1)} = \mathbf{R}^{(t)} + \delta R \frac{\mathbf{R_f} - \mathbf{R}^{(t)}}{|\mathbf{R_f} - \mathbf{R}^{(t)}|} . \tag{3.1}$$

These trial locations are accepted as long as they remain within the allowed parts of configuration space

$$V\left(\mathbf{R}_0^{(t+1)}\right) \leq E_L \quad \Rightarrow \quad \mathbf{R}^{(t+1)} = \mathbf{R}_0^{(t+1)} , \tag{3.2}$$

giving us the straight-line parts of the paths. When a trial step takes the path into a forbidden region, the path is rerouted towards the boundary by moving along the potential-gradient (steepest-descent) path using a Newton-Raphson root search[80] to find the nearest solution of $V(\mathbf{R}) = E_L$.

$$\mathbf{R}_{n+1}^{(t+1)} = \mathbf{R}_n^{(t+1)} - \frac{V\left(\mathbf{R}_n^{(t+1)}\right) - E_L}{\left|\nabla V\left(\mathbf{R}_n^{(t+1)}\right)\right|^2} \nabla V\left(\mathbf{R}_n^{(t+1)}\right) , \quad (n = 0, 1, \ldots) \tag{3.3}$$

The converged answer is then taken to be $\mathbf{R}^{(t+1)}$, the location of the (t+1)-st point on the path. (In practice, the net step length obtained in this way is invariably small enough to produce an effectively continuous path, at least for the kinds of many-body applications we have been pursuing. In particular, our checks show that it routinely satisfies the condition $\left|\mathbf{R}^{(t+1)} - \mathbf{R}^{(t)}\right| < \delta R$.)

This process of stepping directly towards the final point and burrowing out from under any regions with too high a potential, Eqs. (3.1)-(3.3), is repeated until a point on the path lies within $\delta R$ of $\mathbf{R}_f$. The contour length of the path is then computed from the sum[81]

$$\ell = \sum_{t=0}^{t=P} \left|\mathbf{R}^{(t+1)} - \mathbf{R}^{(t)}\right| \quad , \quad (\mathbf{R}^{(0)} = \mathbf{R}_i \, , \, \mathbf{R}^{(P+1)} = \mathbf{R}_f) \quad . \tag{3.4}$$

By construction, the sequence of points $\{\mathbf{R}^{(t)} \, ; \, t = 0, \ldots, P\}$ automatically constitutes a successful path in what we have called the Kuhn-Tucker class, but, as we have noted, such a sequence need not be a true geodesic. To finish our task, we therefore need to optimize the path. We do so by searching for the shortest path lying within a local neighborhood of our unoptimized path, something we can look for with a simple Monte Carlo scheme:

(1) We select a random step t along the path and adjust the configuration $\mathbf{R}^{(t)}$ to some nearby $\mathbf{R}'^{(t)}$ by moving a randomly chosen atom. We then generate a path passing through the configuration $\mathbf{R}'^{(t)}$ by re-applying the path-finding algorithm we have just detailed to the problem of finding two separate partial-paths

$$\mathbf{R}_i \to \mathbf{R}'^{(t)} \quad , \quad \mathbf{R}'^{(t)} \to \mathbf{R}_f \quad .$$

The union of these two segments is deemed the trial path and we calculate its length by computing the sum of the two partial-path lengths using Eq. (3.4). If this trial path is shorter than the length of our current candidate for the geodesic path, we accept the trial move and consider the whole revised set of configurations $\{\mathbf{R}'^{(t)}; t = 0, \ldots, P\}$ to be our new geodesic candidate. If not, we continue to take our previous path to be the correct one.

(2) We repeat step (1), using each new path as the starting point for generating a new trial path. The process is continued until the path length converges. Our final, optimized, geodesic path is then taken to be the sequence of configurations $\{\mathbf{R}'^{(t)}; t = 0, \ldots, P\}$ with the lowest contour length.

    We illustrate the basic geometry of geodesic paths, as well as the distinction between these path and more conventional transition paths, by applying our algorithms to the Müller-Brown potential,[82] one of the standard two-dimensional potential surfaces used to test transition-path finding algorithms (Fig. 4).[4,46,72,83,84] As is clear from Fig. 4a, geodesic paths for a given landscape energy are usually going to be quite different from the minimum→saddle→minimum sequences that make up standard transition paths, in part because the initial and final configurations of our geodesics will rarely coincide with the inherent structures (minima) of the potential surface. The distinctions between geodesics and transition paths, though, go beyond the differences in endpoints. Even if we were to look for the geodesics whose endpoints were located at the potential minima (Fig. 4b), we could find the answers might depend strongly on the landscape energy,[85] and some or all of these answers could be noticeably different from the reaction path.

The particular example we show in Fig. 4b also illustrates the way our algorithms work to find geodesic paths. The original path found by the deterministic part of our algorithm, Eqs. (3.1)-(3.3), does indeed consist of straight lines joined by a segment along the $E_L = V$ boundary curve. Our subsequent Monte Carlo optimization procedure simply minimizes the fraction of the path that lies along the boundary.

It is worth emphasizing that neither our algorithm for finding a Kuhn-Tucker path between our end points nor our scheme for optimizing that path are necessarily the best we can do. In the approach we have been describing, we find the Kuhn-Tucker path by starting at one end point and heading directly for the other (a "drag" method in the language of the field).[41] However, the paths obtained in this fashion will, in general, be different from those constructed by starting from the second end point and heading for the first. That not only allows one to test the putative geodesic paths by seeing if the latter paths are shorter (which does not seem to be the case in our applications), it suggests examining algorithms which employ both end points simultaneously. Our experience to date has been best with drag methods, but it is possible that some more symmetrically double-ended boundary-condition approach[73] might ultimately prove to be superior. It is also quite possible that one can improve on our particular drag method by loosening the restriction that the path head directly for its destination. One might want to explore directions within a narrow (hyper) cone about the direct path, for example.

Generalizations similar in spirit to these are also available for the optimization part of our algorithm. Our Monte Carlo optimization is fundamentally a Metropolis random walk in the space of paths[40,42,43] using changes in path length $\ell$ as a criterion for guiding the walk from path to path (instead of using changes in potential energy from configuration to configuration). Our scheme only accepts trial paths when they lower $\ell$, but doing so simply means that we are effectively taking the "temperature" conjugate to $\ell$ to be zero. When viewed in this fashion, it is easy to see that other approaches might have advantages. We could, for example, carry out the optimization as a simulated annealing process, starting at some finite temperature that we gradually lower to zero. More generally, we could imagine abandoning random walks altogether in favor of a deterministic approach. The key point is that whatever approach one takes to either half of the geodesic-finding problem has to be at least as well suited to the exceptionally high dimensionality of our configuration space as our current formulations are. As we show in Sec. V, the algorithms presented here seem to work reasonably well in the largest examples we present in this paper – those with N = 500 atoms, which require finding the geodesics in a 3N = 1500-dimensional space.

## IV. Models and methods

To help us showcase some of the features of geodesics in liquids, we consider the same two literature models we employed in our companion paper: a single-component atomic liquid,[86] which we expect to be emblematic of ordinary simple-liquid behavior, and the Kob-Andersen binary atomic mixture,[87] a standard example of glass-forming liquid.[88,89] Both systems have pair potentials of the Lennard-Jones form:

$$u_{\alpha\beta}(r) = \begin{cases} u_{\alpha\beta}^{LJ}(r) - u_{\alpha\beta}^{trunc}(r) &, \quad r < r_c = 2.5\,\sigma_{\alpha\beta} \\ 0 &, \quad r > r_c = 2.5\,\sigma_{\alpha\beta} \end{cases}$$

$$u_{\alpha\beta}^{LJ}(r) = 4\varepsilon_{\alpha\beta}\left[\left(\frac{\sigma_{\alpha\beta}}{r}\right)^{12} - \left(\frac{\sigma_{\alpha\beta}}{r}\right)^{6}\right]$$

for each atom of species $\alpha$ separated from an atom of species $\beta$ by a distance r, but with slightly different truncation potentials

*single-component* $\quad u_{\alpha\beta}^{trunc}(r) = u_{\alpha\beta}^{LJ}(r_c) + u_{\alpha\beta}^{\prime LJ}(r_c)(r - r_c)$

*Kob-Andersen* $\quad u_{\alpha\beta}^{trunc}(r) = u_{\alpha\beta}^{LJ}(r_c)$ .

As before, we study the single-component liquid at reduced density $\rho\sigma^3 = 1.058$ with both N = 256 and 500 atoms. The Kob-Andersen mixture, which has potential parameters

$$\varepsilon_{AA} = \varepsilon, \quad \varepsilon_{BB} = 0.5\,\varepsilon, \quad \varepsilon_{AB} = 1.5\,\varepsilon$$

$$\sigma_{AA} = \sigma, \quad \sigma_{BB} = 0.88\,\sigma, \quad \sigma_{AB} = 0.8\,\sigma \quad,$$

is studied at a total reduced density of $\rho\sigma^3 = 1.2$. For this system we also work with both N = 256 total atoms (a mixture of 205 A atoms and 51 B atoms) and N = 500 total atoms (400 A and 100 B).

We find the geodesics of the potential landscapes for these liquids by using the approach outlined in the previous section. But to implement this approach, we need first to generate an appropriate set of configurations that can serve as the endpoints. That is, for each landscape energy $E_L$ we need to pick out statistically independent pairs of configurations separated from one another in the 3N-dimensional configuration space by a prescribed Euclidean distance R. The way we achieve this goal is to sample the energy landscape ensemble by the same Monte Carlo procedure we used in our previous paper:[52] We use a (zero-temperature) Metropolis sampling in which every trial configuration **R** for which $V(\mathbf{R}) \leq E_L$ is accepted and every **R** for which $V(\mathbf{R}) > E_L$ is rejected. The Monte Carlo moves themselves are implemented by randomly selecting an individual atom and attempting to displace it, in each of the Cartesian directions, by a distance δr. (The magnitude of that displacement, in turn, was determined for each $E_L$ by requiring that it produce an acceptance probability of 0.5 in separate calculations.)[90] A complete Monte Carlo step for our N-atom systems consisted of N such attempted moves.

In practice, we generated 99 configuration pairs for each R and $E_L$: Configurations selected from the ensemble were used to generate 11 independent Markov chains, with each chain supplying 9 configuration pairs. Those pairs, in turn, were produced by assigning the starting configuration of each chain, $\mathbf{R}_1^i$, to be the first endpoint and propagating the chain until it reached a configuration $\mathbf{R}_1^f$ separated by R from the initial configuration. We designated the pair ($\mathbf{R}_1^i, \mathbf{R}_1^f$) to be the first pair of endpoints. We then propagated 1000 additional Monte Carlo steps and assigned the resulting configuration $\mathbf{R}_2^i$ to be the first endpoint of the second pair. This procedure was repeated until we accumulated the desired number of configuration pairs.

The equilibrium equivalence of the canonical and landscape ensembles actually allows us to use a molecular dynamics approach to finding the geodesic endpoints if we care to. We therefore implemented an equivalent molecular dynamics procedure by starting with configurations selected from thermally equilibrated systems at a range of temperatures, running a collection of 11 independent NVE classical trajectories for each temperature and selecting 9 configuration pairs from each trajectory. As with the Monte Carlo approach we selected the endpoints from our trajectories by first adopting a configuration and propagating until we reached another configuration at the correct distance, then running for 1000 additional time steps $\delta t$ and repeating the procedure. The precise value of the landscape energy $E_L$ associated with each pair of configurations ($\mathbf{R_i}$, $\mathbf{R_f}$) was taken to be the greater of ($V(\mathbf{R_i})$, $V(\mathbf{R_f})$). Our molecular dynamics calculations for this process and for our subsequent molecular-dynamics-based diffusion constant calculations were carried out using the velocity-Verlet algorithm,[91] adopting a time step $\delta t = 0.001\ \tau_{LJ}$ for the single component liquid and $\delta t = 0.001\ \tau_{LJ}$ ($k_BT/\varepsilon > 1$), $\delta t = 0.003$ $\tau_{LJ}$ ($k_BT/\varepsilon < 1$) for the binary liquid. (As usual, $\tau_{LJ} = \left(m\sigma^2/\varepsilon\right)^{1/2}$, with the atomic mass m assumed to be the same for all of our atoms).

The geodesic endpoints used in creating the graphs shown in this paper are all derived from this latter, molecular-dynamics-based, approach. However, explicit comparison of the geodesic lengths resulting from the Monte-Carlo and molecular dynamics approaches (not shown) reveals that the averages $\langle (R/g)^2 \rangle$ are identical within the error bars of each method. This consistency offers a welcome measure of reassurance regarding the reliability of our geodesic-finding algorithm. More than that though, in the absence of such a check, one might have wondered whether obtaining endpoints from classical trajectories was somehow building dynamical-path information into what should be a purely geometrical analysis of the potential-energy landscape. Given that we do obtain the same results from the endpoints of actual trajectories and from endpoints separated by Monte Carlo paths with contours $10^3$-$10^4$ times longer than the geodesic paths, it seems unlikely that there is any such dynamical bias in our choice of endpoints.

As a final point, we note that we will be reporting all of our results as functions of temperature T, even though all of our geodesic calculations are carried out as a function of landscape energy $E_L$. We can do so, because Eqs. (1.1) and (1.2) can be used to translate one variable into the other. The precise relationships between landscape energy and temperature for the two model liquids we examine in the next section were worked out in our previous paper[52] and are summarized in Fig. 5.

## V. Numerical results for geodesic paths in liquids

Since we know the relationship between temperature and landscape energy, we can begin to study how the lengths of the geodesic paths g change with the thermodynamic state of our liquids. We have applied the algorithm described in Sec. III to compute these lengths as a function of $E_L$ and used Fig. 5 to convert these findings into plots as a function of temperature. The results are displayed in Figs. 6 and 7.

It is clear from both of these graphs that once we examine long enough geodesics, the lengths of the geodesics g scale with the Euclidean distance R between the endpoints. That is, the ratio of the two (R/g) is what is the fundamental property of a liquid. Figures 6 and 7 show that this ratio converges to an R- and N-independent value at each $E_L$ (or T). The value of R needed to reach the asymptotic regime seems to scale with $N^{1/2}$ (meaning that to compare different N values, we need to keep the value of $R^2/N$ constant) but that is sensible; since R is a distance in the 3N-dimensional configuration space of the system, the linear distance sampled by any one molecule is only of the order of $R/(N)^{1/2}$. Once we are in the asymptotic regime, though, the details of endpoint separation and system size become irrelevant.

So what happens to these geodesics as one lowers the temperature? As $E_L$ (or T) decreases, the ratio (R/g) decreases noticeably, confirming the idea that the paths are becoming progressively more convoluted at lower T. In our "barrier-free" language, we would say that it is becoming harder and harder to find efficient paths between configurations as more and more of the shortest routes are closed off. Notice, though, how much more pronounced this phenomenon is for the glass-forming binary system than it is for the ordinary liquid. The square of the ratio $(R/g)^2$ drops by an order of magnitude on lowering the temperature in the glass-forming system, whereas it only drops by a factor of 2 in the ordinary liquid.

As we noted in Sec. II, this very ratio is precisely what we need in order to compute the diffusion constants of these liquids. All we need to do to arrive at explicit landscape predictions for the diffusion constants is to have enough dynamical information to set the overall scale. In accordance with our previous discussion, we therefore fit high-temperature molecular dynamics results to Eq. (2.12) to determine the mobility constant μ for each system and plotted the diffusion constants predicted by Eq. (2.11) in Fig. 8. We then compared these results with actual dynamical values obtained by numerically integrating the velocity autocorrelation functions derived from molecular dynamics runs on these same systems.

The results for the ordinary liquids do little more than confirm that, over the liquid range, Eq. (2.12) is itself a fairly accurate representation of the temperature dependence of typical (constant-density) liquid diffusion constants. However the agreement between the exact dynamics and the non-dynamical, purely geometric, predictions of the landscape theory is reasonably impressive for the glass-forming binary system. The diffusion constants of both components undergo a two-order-of-magnitude drop over the temperature range shown in the figure, a drop apparently explained quantitatively by the parallel growth in the lengths of the geodesic paths over the corresponding range of landscape energies.

As has been discussed in some depth in the literature,[87-89] the precipitous decline in diffusion constants for this system coincides with its mode-coupling transition.[92] So, what we are finding is that the landscape results are essentially predicting the location of that transition – without any need for any mode-coupling-like dynamical arguments. Moreover, in our companion paper we put forth the argument that the landscape approach offered a possible geometrical interpretation of this phenomenon: a percolation transition as a function of landscape energy.[52] We suggested that when the potential energy dropped below $E_c$, the critical potential energy associated with the mode coupling temperature $T_c$, the landscape suddenly separated into a macroscopically significant number of disconnected regions. Any classical trajectories trying to traverse this landscape with a total energy less than $E_c$ would therefore have to be nonergodic.

The evidence for our interpretation in our previous paper[52] was largely thermodynamic: Landscape energy values below $E_c$ lost their association with a unique value of configurational temperature. We noted, in addition, that the observation of a "sharp change in local topography" at this point by Sastry, Debenedetti, and Stillinger[12] had its analog in our own finding that the disconnected regions of the potential surface had the thermodynamic characteristics of localized harmonic wells. To these points, we can add one piece of dynamical evidence from the current work. The energy $E_c$ marks a threshold below which we were unable to find endpoints separated by distances large enough to make the geodesic lengths relevant to diffusion. Below $E_c$ neither the molecular dynamics nor the Monte Carlo methods we used (Sec. IV) succeeded in generating configurations far enough apart for Eq. (2.11) to apply – a somewhat surprising result given that we might have expected that it would be our geodesic path finding algorithm that would prove to be the ultimate limitation to approaching the glass transition.

We finish our presentation of the numerical results by noting that since we have actual geodesic paths in hand, it is possible to confirm one of our key assumptions about them: that the potential energy really is close to being constant along such paths. Although our ensemble imposes restrictions only on the *maximum* potential energy the system can have as it traces a path, the results displayed in Fig. 9 show that the potential energy does indeed remain constant as one moves along the geodesics of the Kob-Andersen liquid. Of course, part of this finding is not all that surprising: The segments of these paths that lie along the $V(\mathbf{R}) = E_L$ boundaries are obviously going to be constant potential energy contours. What might be more surprising, given the picture of these paths suggested by Fig. 2, is the constancy of the potential energy for the segments between those boundaries. This result is purely a consequence of the high dimensionality of configuration space. For large d, the vast majority of the configurations in any d-dimensional closed volume lie within a small distance of the surface – not within the interior – and the surface, for us is $V(\mathbf{R}) \approx E_L$.[93]

## VI. Concluding remarks

Our central interest here is in understanding what it is about the potential energy landscapes of slow condensed-matter problems that dictates the most efficient possible motion when the dynamics is slow. In some cases answering this question is a matter of finding a transition state between two well-defined potential minima (a process that can be hard enough in a high-dimensional space), but more generally it means discerning the fastest routes for our system to travel the long distances from one arbitrary configuration to another – whether the process involves any stationary points or not.

To help us think about this more general pathway problem we opted to ask an even simpler question: what route would our system follow if it wanted to cover what is, in some sense, the least distance between two specified locations? Even asking this question required a shift in perspective. Strictly speaking, the shortest route in the canonical ensemble is always a straight line between the endpoints; there are no intervening barriers so high that there would not be at least some probability of climbing over them. However, within the potential energy landscape ensemble this same question is a bit more meaningful. When diffusion constants are small, we have argued that the shortest (geodesic) paths in this ensemble are representative of the actual paths the system takes. The slowest diffusion therefore occurs when the geodesic pathways through the landscape becomes so tortuous that they become significantly longer than the direct distance between the endpoints.

What we have demonstrated in this paper is that all of these notions are well-defined and computationally accessible, even when one has the rugged, high-dimensional energy landscapes characteristic of supercooled liquids. For one thing, geodesic paths in the potential energy landscape ensemble have specific geometric features that help us locate them. The paths follow the shortest barrier-skirting routes they can, so in atomic systems, for example, the paths have to be constant-potential-energy contours connected by straight-line segments. In addition, the geodesics have finite-size scaling properties that, along with other checks, can be used to help us test our path finding procedures.

The real import of this geodesic formulation is that it demonstrates that there is critical information about the long-time dynamics that can be extracted directly from the geometry of the potential landscape. The fact that the mass of our atoms, for example, does not even enter our diffusion constant relationship, Eq. (2.11) (other than in an overall constant), is testimony to the intrinsically non-dynamical character of this approach. The fact that we can use landscape properties to understand the decline in diffusion constants as we approach a mode coupling transition is a similarly striking piece of evidence. The quantitative accuracy of the formalism's predictions for these diffusion constants is notable as well, but even leaving that aside, the suggestion from these results that there is a time-scale-independent percolation transition underlying such manifestly time-scale-dependent properties of glass-forming materials[94] is a classic example of the kind of hypothesis that our approach is designed to examine. We should be able to achieve a more detailed characterization of any such transition just by quantifying additional geometrical features of potential surfaces — without having to work backwards from time correlation functions or transport coefficients.

Having made these points, we should also emphasize that the formalism has no current predictions for dynamics below the mode coupling transition. One can still measure the lengths of geodesics below this transition, but it is not obvious that these, relatively short, lengths should control the ultimate rate of transport throughout the system. Whether there are other properties of the landscape topology that start to play an equivalent role at these low landscape energies is yet to be determined. It is clear, however, that there are many characteristics of our geodesics that one can look at besides their lengths. Because the special molecular motions involved in the geodesic path for a given process have to include whatever mechanical events there are that are critical to that process, it should be possible to interrogate the geodesic paths to find the molecular mechanisms behind many different slow processes. If molecular rotation does contribute differently from center-of-mass translation[28] or if string-like motions are important,[95] the geodesic paths should highlight these features. Equally interestingly, if one thinks that the process of hard-sphere-like jamming[96] is really representative of what occurs with more realistic intermolecular potentials, one should be able to make detailed comparisons of the onset of slow dynamics for these two kinds of systems. Geodesics in the energy landscape ensemble should be equally well defined and equally accessible for hard-core liquids.

**Acknowledgements:** We thank Kenneth Schweizer, David Wales, and Guohua Tao for helpful discussions. This work was supported by NSF grant nos. CHE-0518169 and CHE-0131114.

**Figure Captions**

**Figure 1**: The potential energy landscape ensemble. Both panels show the potential energy V as a function of the configuration **R** of the system with the white regions (those lying below the landscape energy $E_L$) the allowed regions and the shaded regions the forbidden regions. (Note the different convention from Fig. 1 in Ref. 52) The lower panel redraws the upper panel as a contour plot, emphasizing the complex, multiply-connected, many-dimensional character of the allowed regions.

**Figure 2**: Geodesics in the potential energy landscape ensemble. The upper panel shows a possible shortest path between initial and final configurations *i* and *f*. The path consists of straight-line segments combined with segments skirting the boundary of the forbidden regions, as required by the Kuhn-Tucker theorem. The lower panel shows (as dashed lines) some alternative possibilities for path segments that also obey the Kuhn Tucker theorem.

**Figure 3**: Our algorithm for finding Kuhn-Tucker candidates for geodesic paths. Panel **(a)** shows how one starts at the initial configuration *i* and heads directly for the final configuration *f* (dark solid line), a procedure that eventually leads to a slight incursion (dotted lines) into a forbidden region, shown here as a circle. This misstep is remedied by following the gradient of the potential until one returns to the boundary of the forbidden region (the black dot). The whole process is then repeated, as show in panel **(b)**. The final path found by successive applications of the process is shown in **(c)**.

**Figure 4**: Different pathways through the two-dimensional Müller-Brown potential energy landscape. The top panel contrasts an energy-landscape geodesic path (for landscape energy $E_L$ = -0.20) with the reaction path. The geodesic path goes between two arbitrarily chosen points A and B and stays reasonably close to the E = -0.20 constant-potential-energy contour. The reaction path travels from one minimum (the "reactants" R) to another (the "products" P) by way of yet another minimum (located at (-0.05, 0.47)) and two different saddle points, (-0.82, 0.62) and (0.22, 0.30). The bottom panel compares the reaction path with a number of different energy landscape paths between R and P. The open stars denote the geodesic candidate predicted by our (unoptimized) Kuhn-Tucker-path finding algorithm. The path starts at R and heads directly for P, turning left when it encounters a barrier, and then resumes its direct path to P as soon as it clears the barrier. The filled stars show a shorter path-length geodesic candidate produced by an intermediate level of optimization, and the open circles show the final geodesic path – which differs noticeably from the reaction path, especially in the product region, largely because of the way that the reaction path has to pass through the left saddle point.

**Figure 5**: Energy/temperature relations in the landscape and canonical ensembles for the two different model liquids studied in this paper. Both panels show the configurational temperature from Eq. (1.2) plotted as a function of landscape energy $E_L$ (solid lines with barely visible vertical error bars) and compare that to the canonical average potential energy $\langle E \rangle$ plotted versus canonical temperature (squares with horizontal error bars). The hatched region in the upper panel is the solid/liquid coexistence region and the dashed grey line in the lower panel is a literature fit to canonical ensemble data for the Kob-Andersen liquid (Ref. 88). Given the scale of the figure, the only place where the reader can probably see an error bar for the landscape results is within the coexistence region of the upper panel. Both panels are for systems with N = 256 atoms.

**Figure 6**: Geodesic path lengths as a function of configurational temperature for the two model liquids studied in this paper. This figure shows how the ratio of R, the Euclidean path length (the direct end-to-end distance) to g, the geodesic path length, converges as the paths become longer. Both panels are for systems with N = 256 atoms.

**Figure 7**: Geodesic path lengths as a function of configurational temperature for the two model liquids studied in this paper. As with Fig. 6, this figure looks at how the square of the ratio to Euclidean to geodesic path length, $(R/g)^2$ decreases with decreasing temperature, but here the emphasis is on the invariance of the results to N, the number of atoms, for a given $R/N^{\frac{1}{2}}$ ratio.

**Figure 8:** Reduced diffusion constants $D^* = D \sqrt{m/\varepsilon\sigma^2}$ for our two liquids as a function of temperature T. Molecular-dynamics derived diffusion constants (points) are compared with landscape geodesic predictions (lines) for the single-component atomic system, **(a)**, and for both the A and B particles in the Kob-Andersen binary case, **(b)** and **(c)**, (with Fig. 5 used to translate $E_L$ into T). The location of the literature mode-coupling transition (Refs. 87, 89) is indicated by the arrow in **(c)**.

**Figure 9**: The variation of potential energy along typical geodesic paths in the Kob-Andersen liquid. Shown here are results for 9 different geodesics obtained from the 9 different configurational temperatures ($T^* = k_BT/\varepsilon$) shown in the key. The distance traveled along each path is denoted by s. The sudden jumps and dips at the endpoints are related to the fact that the potential energies of the beginning and ending configurations are often different from each other and from the landscape energy.

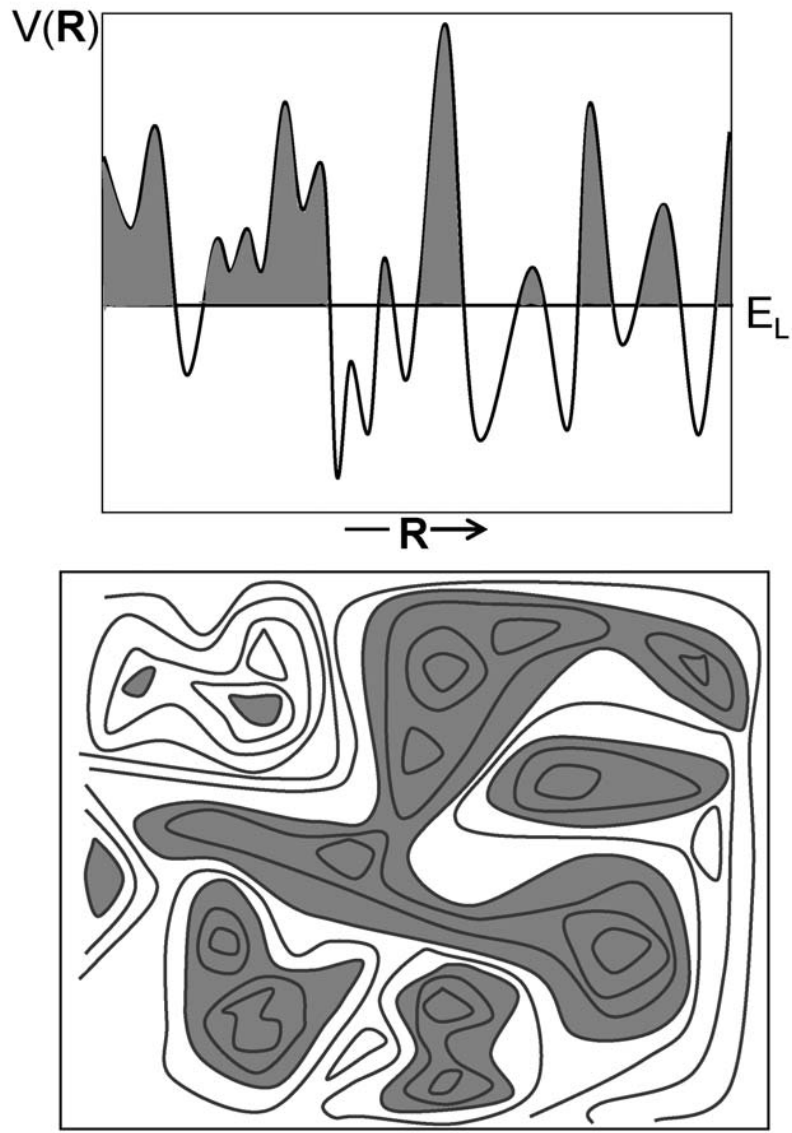

Figure 1

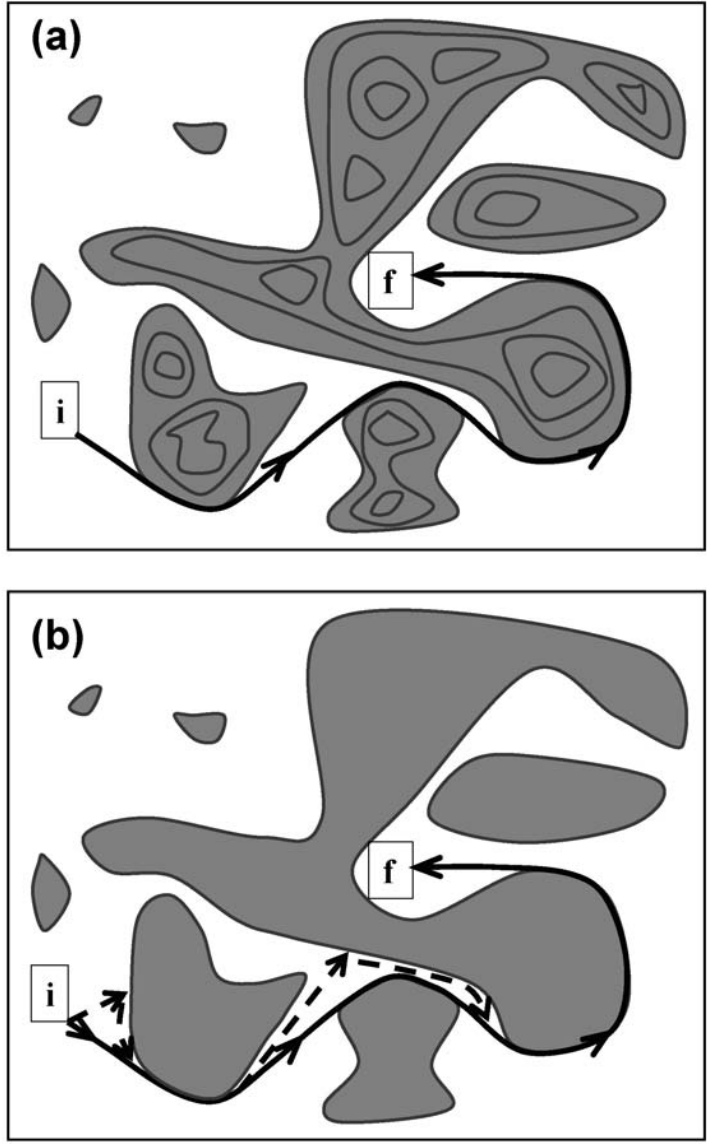

Figure 2

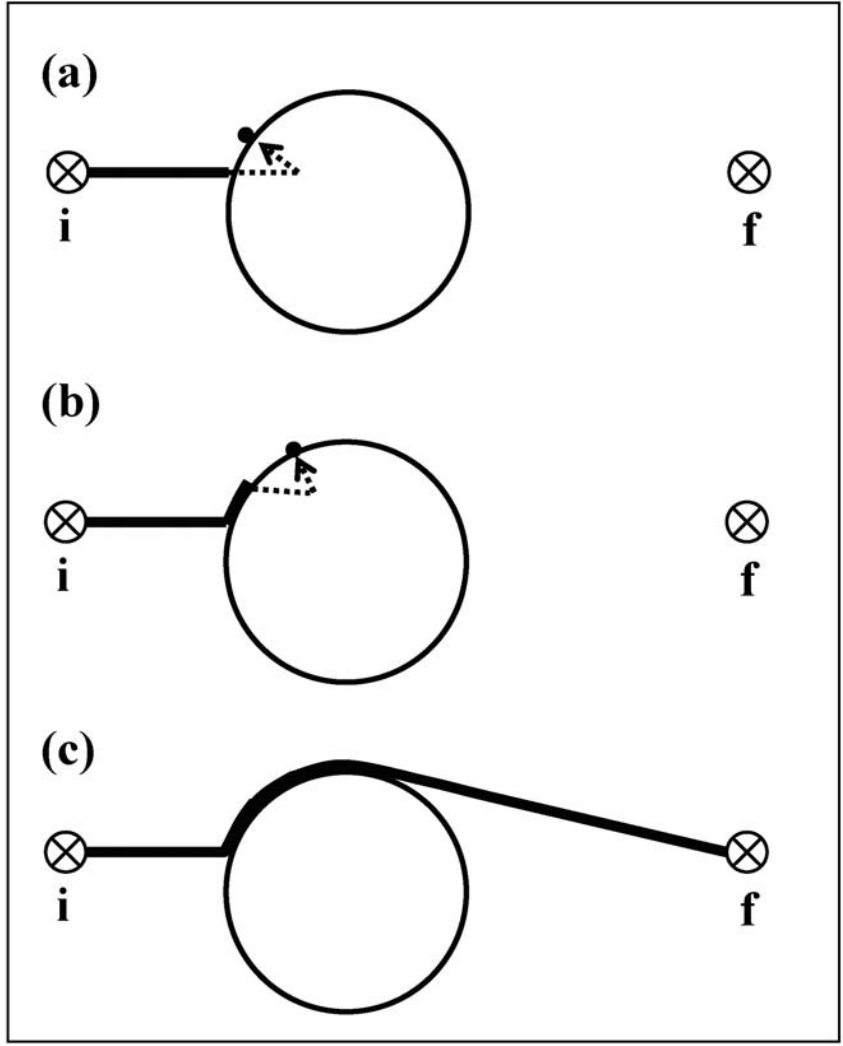

Figure 3

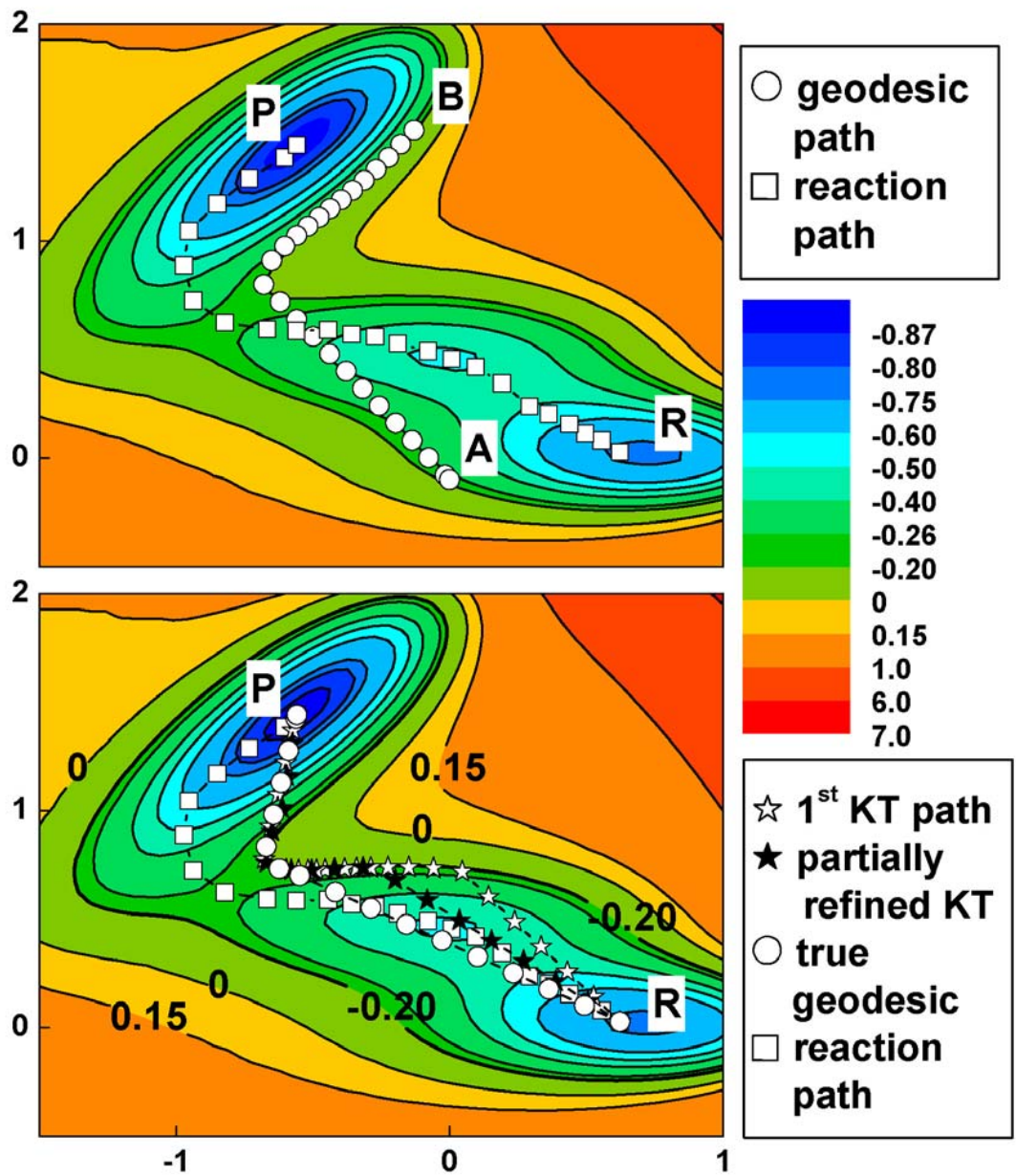

Figure 4

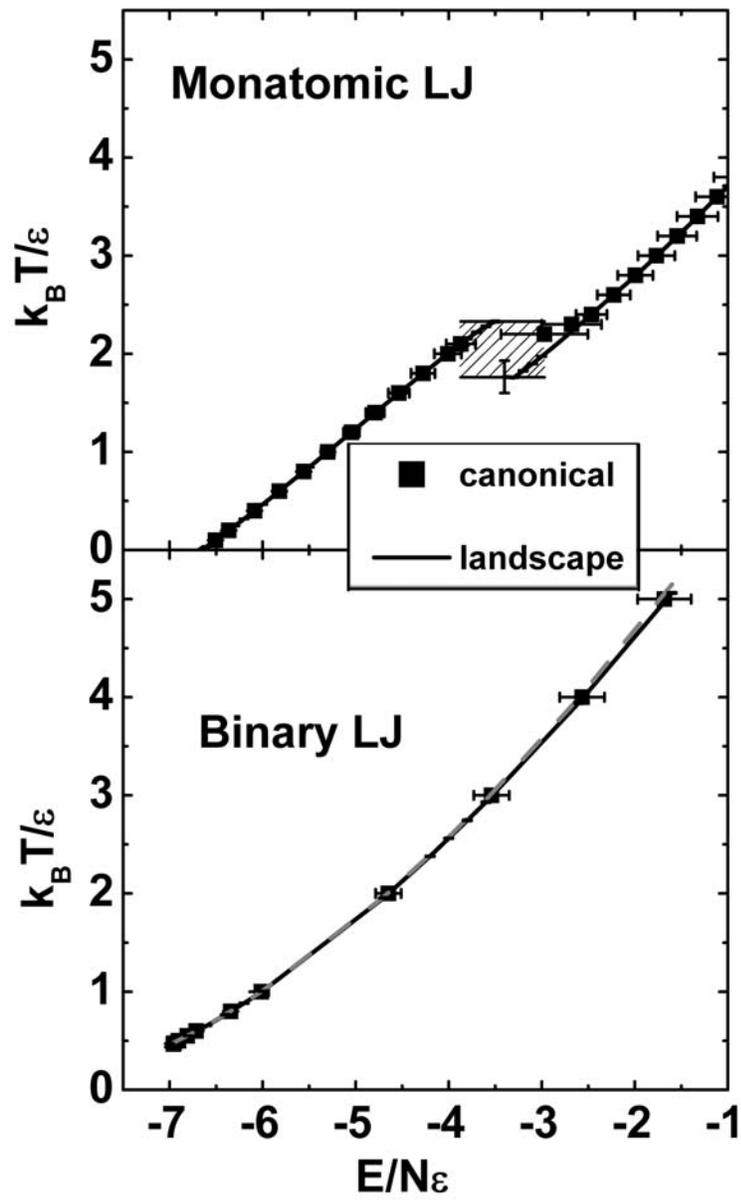

Figure 5

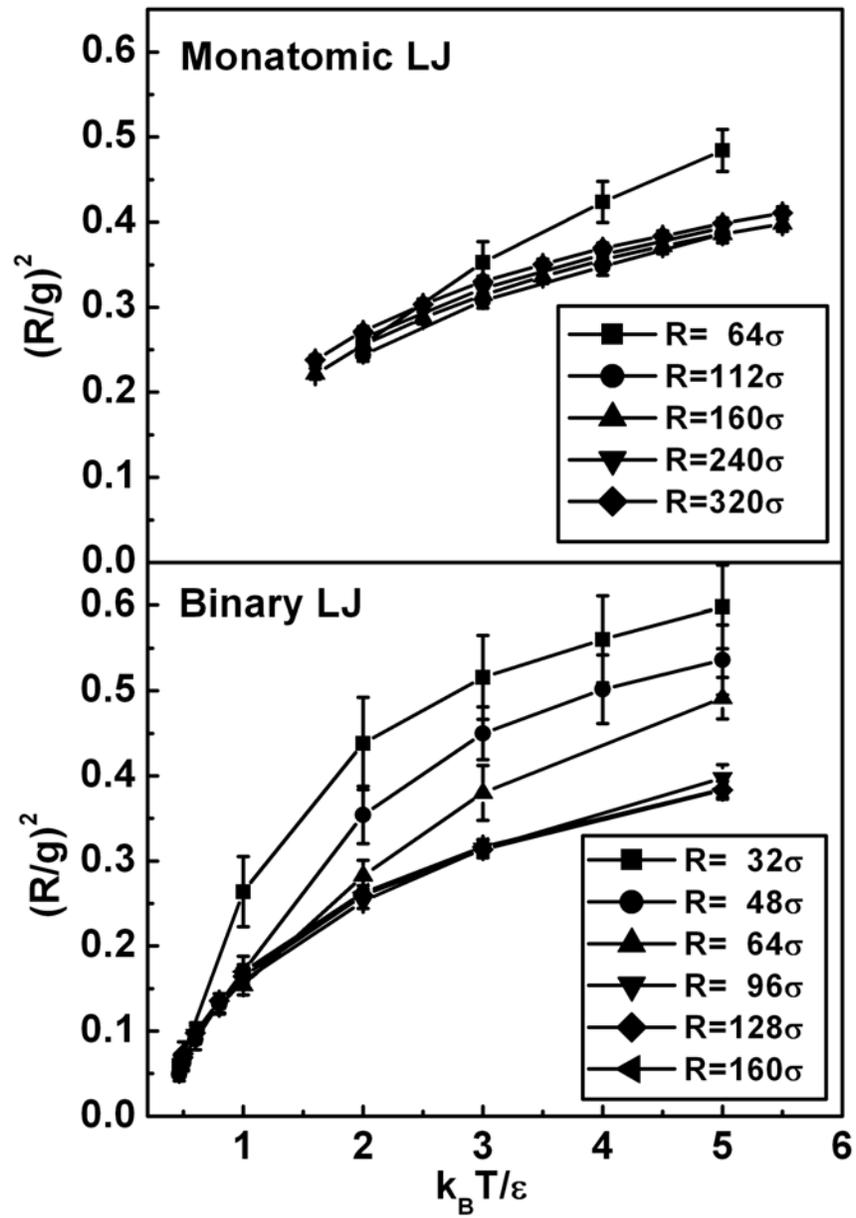

Figure 6

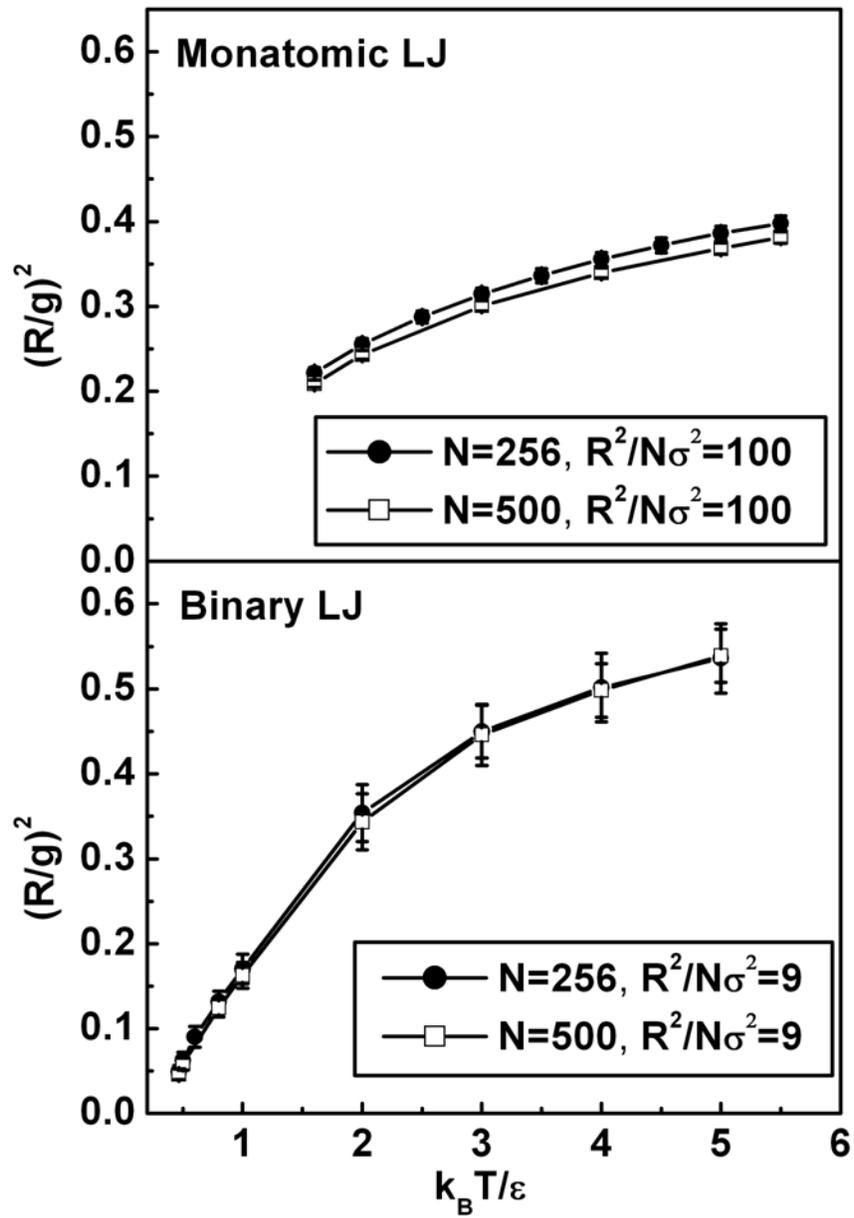

Figure 7

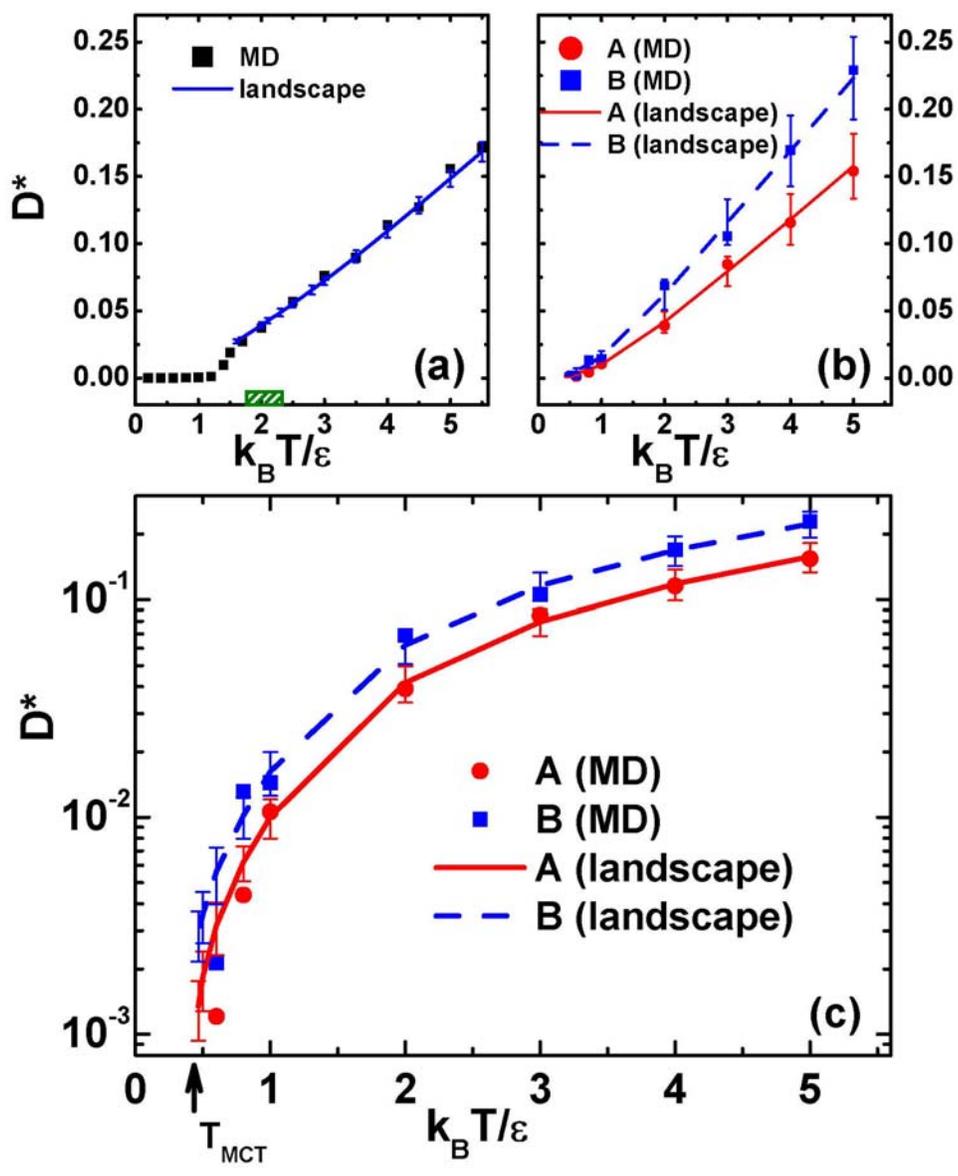

Figure 8

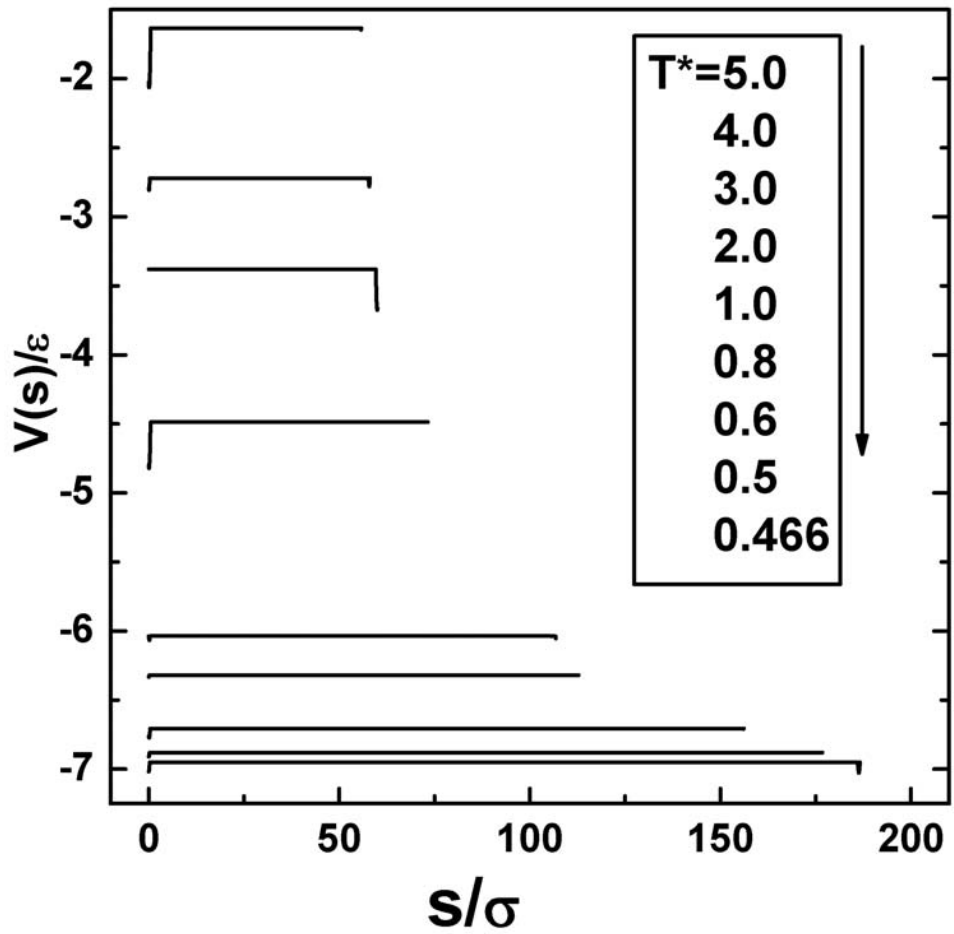

Figure 9